\documentclass[10pt,letterpaper]{article}

\usepackage{cogsci}

\cogscifinalcopy 

\usepackage{graphicx}
\usepackage{pslatex}
\usepackage{apacite}
\usepackage{float} 

\title{The Dynamics of Collective Creativity in Human-AI Hybrid Societies}
 
\author{
\large
\textbf{Shota Shiiku\textsuperscript{1, 5}, Raja Marjieh\textsuperscript{2}, Manuel Anglada-Tort\textsuperscript{3, 5}, Nori Jacoby\textsuperscript{4, 5}} \\
\textsuperscript{1}Graduate School of Science and Technology, Shizuoka University, Hamamatsu, Japan\\
\textsuperscript{2}Department of Psychology, Princeton University, Princeton, USA\\
\textsuperscript{3}Department of Psychology, Goldsmiths, University of London, London, UK\\
\textsuperscript{4}Department of Psychology, Cornell University, Ithaca, USA \\
\textsuperscript{5}Max Planck Institute for Empirical Aesthetics, Frankfurt am Main, Germany
}

\begin{document}

\maketitle

\begin{abstract}
Generative AI is shaping an increasingly hybrid society, where ideas and cultural artefacs are created both by humans and intelligent machines. Human creativity is influenced in complex, nonlinear ways by the actions of AI-driven agents within their social networks, but these influences are difficult to measure using traditional methods. This study examines how human-AI interactions shape the evolution of collective creation within large-scale social network experiments,  where human and AI participants collectively create stories. Participants (either humans or AI) joined 5×5 grid-based networks in which stories were selected, modified, and shared over many iterations. Initially, AI-only networks showed greater creativity (rated by a separate group of human raters) and collective diversity of stories than human-only and human-AI networks. However, over time, hybrid human-AI networks became more diverse in their creations than AI-only networks. In part, this is because AI agents retained little from the original stories, while human-only networks preserved continuity. These findings highlight the value of experimental social networks in understanding human-AI hybrid societies.

\end{abstract}

\section{Introduction}
From cave paintings to symphonies, creativity has defined the human experience – our ability to imagine new possibilities and bring them to life. Today, we stand at a fascinating turning point as generative artificial intelligence (AI) is transforming the creative process by which humans formulate ideas and put them into practice \cite{epstein2023art}. AI models are not merely assisting tools but active partners in the creative process \cite{collins2024building, brinkmann2023machine}, collaborating with writers, musicians, and artists in unprecedented ways. In this hybrid society, networks of interacting humans and intelligent machines constitute complex social systems for which the quality of the collective outcomes cannot be deduced from either human or AI behaviour alone \cite{tsvetkova2024new}. However, studying how collective creativity emerges in hybrid networks of humans and AI agents remains a key challenge.

While generative AI has been shown to increase the quality and efficiency of routine tasks – such as customer support, programming, and academic writing \cite{vaccaro2024combinations}, its impact on human creativity remains poorly understood. Some evidence suggests that human-AI collaboration may enhance individual's creativity \cite{doshi2024generative, lee2024empirical}, increasing the efficiency and speed of generating new ideas. In contrast, at the same time aesthetic and cultural biases embedded in generative AI models can limit global diversity, leading to homogenization effects in art and culture \cite{anderson2024homogenization, doshi2024generative}. This research demonstrates the potential impacts of generative AI on human creativity, but we still know little about how good ideas emerge and evolve within human-AI social networks, and the mechanisms that enable effective collaboration between humans and machines.

Our approach introduces a method that is both ecologically valid and open-ended, enabling the study of large-scale social interactions involving hundreds of participants within social networks. Here, we leveraged our recent capability to design experiments that incorporate real human participants within experimental social networks \cite{marjieh2025characterizing}.
By using iterative storytelling as a creative task, we examine how ideas propagate, transform, and diversify in networks the involve humans, AI, and hybrid human-AI collaborations. This framework allows us to analyze the complex interplay between human and AI agents in dynamic creative processes at a collective level.



\begin{figure*}[t]
\begin{center}
\includegraphics[width=0.8\textwidth]{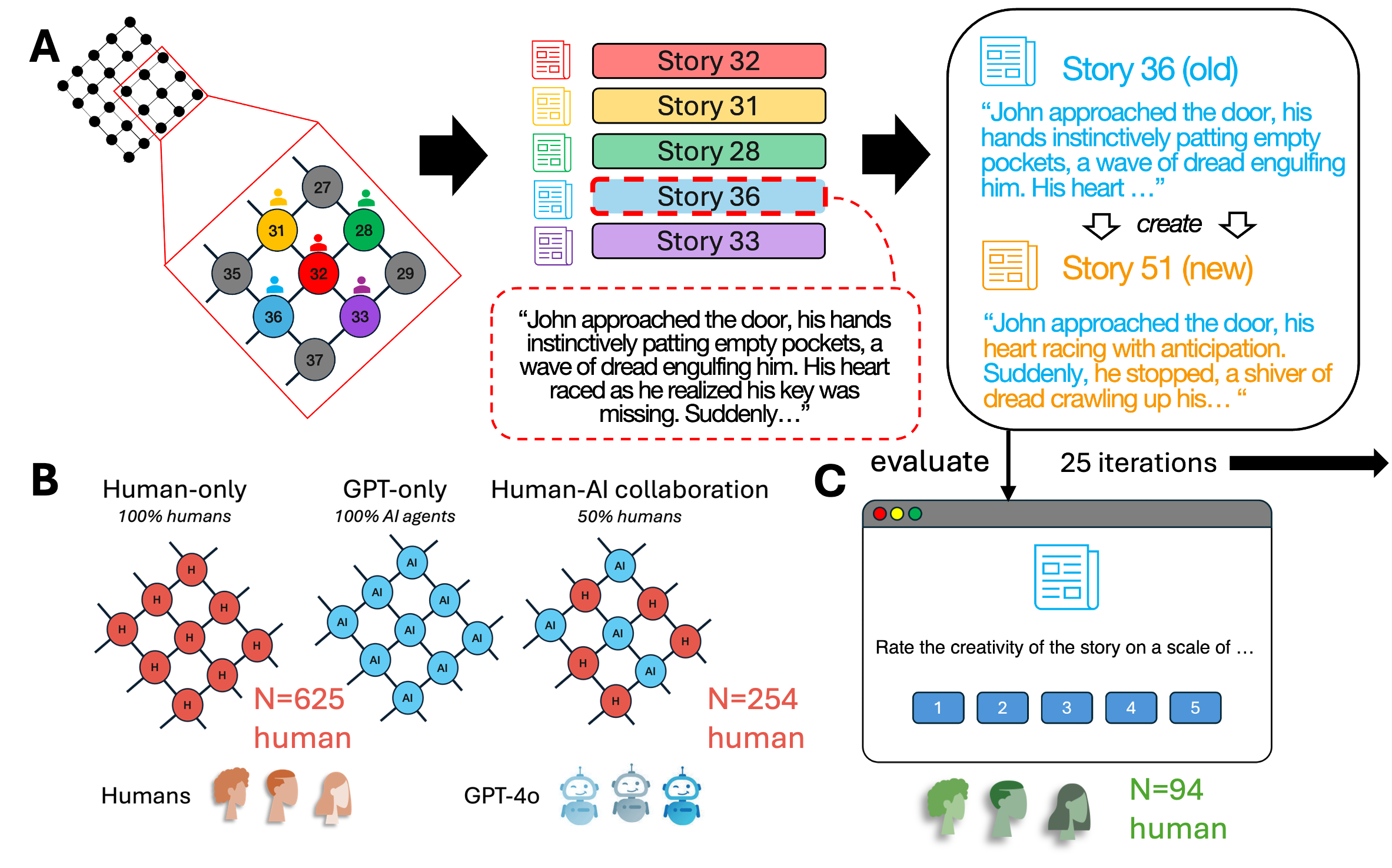}
\caption{Experimental framework for studying collective creativity. (A) Participants join social networks and engage in a creative writing task where short stories are selected, modified, and transmitted over many iterations (B) We study three network configurations: human-only, AI-only, and human-AI. (C) The creativity of stories is assessed by a separate group of human raters.}
\label{over}
\end{center}
\end{figure*} 

\section{Background}
\subsection{Beyond Tools}

Generative AI models can now produce high-quality, human-like content across various modalities, including text (Vaswani et al., 2017; Ouyang et al., 2022), images (Ramesh et al., 2022), and speech (Kumar et al., 2023). This has led to the development of a diverse range of AI-driven tools. 
These include coding assistants like GitHub Copilot, applications in medicine (Jumper et al., 2021), robotics (Murphy, 2019), autonomous vehicles (Badue et al., 2021), and even AI-assisted scientific research (Romera-Paredes, 2024).

Yet, how effective are AI tools? Despite increasing investment in development, research on AI-assisted decision-making presents mixed findings. While some studies indicate that AI enhances performance in domains such as healthcare, customer service, and scientific research \cite{liu2021understanding,chen2023understanding}, others suggest the opposite \cite{bansal2021does, zhang2020effect}. A recent systematic review and meta-analysis of 106 experimental studies (Vaccaro et al., 2024) confirmed this divide, though, on average, human-AI teams performed better than humans working alone.

Rather than viewing AI systems merely as assisting tools, a more promising conceptual shift—drawn from the science of collaborative cognition—is to understand them as "partners in thought": systems designed to be reasonable, insightful, knowledgeable, and trustworthy, capable of actively thinking and creating with humans \cite{collins2024building}. This perspective moves beyond the instrumental view of AI as passive aids and toward recognizing their potential as cognitive collaborators. As AI
models continue to advance and proliferate, they will become increasingly integrated into human interactions and everyday activities, participating in complex, multimodal problem-solving and contributing to collaborative creative processes.

\subsection{AI and Collective Dynamics}

As AI systems become embedded in everyday human activity as collaborative partners, their impact extends beyond individuals to influence collective behavior and social systems at large \cite{brinkmann2023machine, collins2024building,tsvetkova2024new}. Thus, it is becoming incredibly important to understand the role of generative AI in shaping collective behavior, focusing on the emergent structures that arise from interactions between multiple individuals and AI agents simultaneously \cite{tsvetkova2024new}.

Recent advances in computational and experimental techniques now make it possible to study how collective behavior emerges in social networks \cite{centola2022network, malone2022handbook}. Researchers have examined the effects of network size and topology in artificial social networks composed of human agents, revealing how structural factors influence consensus formation and collective intelligence \cite{centola2015spontaneous, derex2013experimental, rand2011dynamic}. For instance, \citeA{shirado2017locally} demonstrated that introducing noise into a system—such as simple bot agents—can enhance consensus by pushing the system out of local minima. However, previous studies have focused almost exclusively on simple, one-dimensional tasks, where solutions are predefined within the problem space, such as problem-solving, inference, and decision-making. Creativity, in contrast, requires engagement with open-ended challenges, vast exploration-exploitation spaces, and complex problem-solving environments. 

\subsection{AI and Collective Creativity}

Creativity is typically assessed along two core dimensions: novelty and usefulness \cite{kaufman2010cambridge, amabile1982social}. Novelty captures the extent to which an idea deviates from existing conventions or expectations, while usefulness reflects the idea’s practicality, coherence, or relevance, such as whether a short story could plausibly be published as a book. Beyond these individual-level criteria, creativity can also be understood at the collective level, where it reflects the diversity of ideas produced across a group or system \cite{maher2012computational,sonicrim2001collective,parjanen2012experiencing}. In this view, a creative ecosystem is not just one that generates isolated novel ideas, but one that fosters a wide range of distinct and meaningful outputs.

Generative AI can impact creativity in at least two ways. AI can enhance creativity by acting as a cognitive catalyst, suggesting unexpected ideas, combining concepts in novel ways, and expanding the exploratory space available to human creators \cite{collins2024building,muller2024group}. Conversely, AI can hinder creativity by promoting convergence and homogenization \cite{bengio2024managing, anderson2024homogenization}. Understanding when and how AI enhances versus constrains collective diversity is thus critical for assessing its long-term impact on human cultural production. \citeA{doshi2024generative} demonstrated that while generative AI can enhance individual creativity, it may simultaneously reduce collective diversity, the range and originality of ideas produced by a group. This study provides an important initial look at the collective consequences of AI-driven creativity, but it did not capture how these effects evolve among diverse agents interacting over time within social networks. 

\subsection{Creativity Research}

Creativity has been studied extensively in psychology and computer science. In psychology, human creativity has been investigated using psychometric measures developed over decades \cite{kaufman2010cambridge}. Classic “creativity tests”—such as the Structure of the Intellect divergent production tasks \cite{guilford1967nature} and the Torrance Tests of Creative Thinking \cite{torrance1966torrance}—remain widely used today. These assessments typically involve language-based tasks, such as listing unusual uses for a common object or generating non-repeating words. More recent advances leverage computational models of semantic networks to address the limitations of these tests. These models allow researchers to quantitatively trace how individuals navigate conceptual spaces, revealing cognitive search strategies associated with creative thought \cite{beaty2023associative}. Beyond verbal tasks, some studies have explored non-linguistic forms of creativity—such as visual art production in grid-based environments—to model how people balance exploration and exploitation in open-ended tasks using computational techniques \cite{kumar2024comparing, hart2017creative, hart2018creative}. However, this research primarily treats creativity at the individual level and does not consider AI as an integral part of the creative process

In computer science, creativity has traditionally been approached from the perspective of computational creativity, a field that explores how machines can be programmed to perform tasks typically associated with human creative behavior \cite{wiggins2018computational, colton2012computational}. This includes the development of algorithms that can generate music, visual art, and poetry. Methods range from rule-based systems and evolutionary algorithms to neural networks and large-scale generative models. A central aim is to build systems that exhibit properties of creativity—such as novelty, value, and surprise \cite{boden2004creative, lamb2018evaluating}. Recent advances in generative AI, particularly large language and image models, have expanded the scope of machine-generated content, making it possible to simulate high-level creative tasks with increasing fluency and realism \cite{ramesh2022hierarchical, brown2020language}. However, this work often emphasizes output quality rather than the cognitive and social processes underlying human-AI creativity, and tends to focus on individual agents rather than collective dynamics. Integrating insights from psychology with computational methods offers a promising path forward for understanding creativity as a collective process that emerges in networks of interacting humans and machines.


\section{Method}
\subsection{Participants}
Participants were recruited online via Prolific\footnote{\url{www.prolific.com}}. All participants were recruited from the UK and identified English as their native language. In total, we recruited 879 human participants and conducted 996 calls to GPT-4o~\cite{hurst2024gpt}, using the version released on September 3, 2024. Particiapnts provided informed consent in accordance with an approved ethics protocol (Max Planck Ethics Council \#202142), and were compensated at a rate of £9 per hour.

\begin{figure*}[t] 
\centering
 \includegraphics[width=\textwidth]{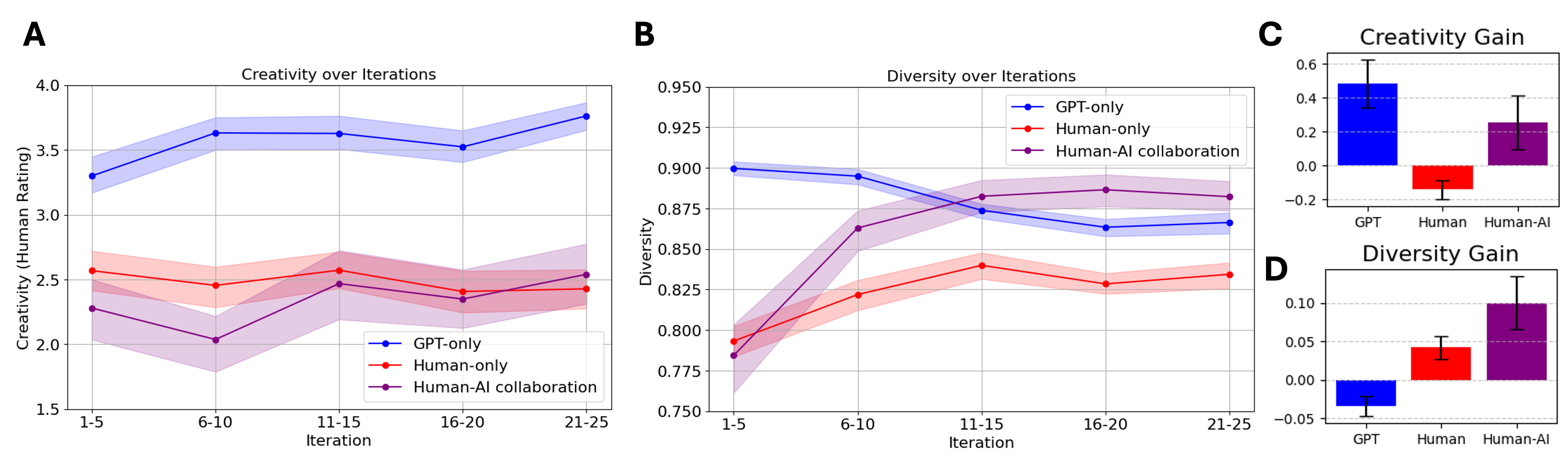} 
 \caption{ The dynamics of collective creativity. (A) Mean creativity ratings of stories over time, as evaluated by human participants.
(B) Diversity of stories (inverse similarity) over time. The horizontal axis represents the 25 iterations, grouped into five sets of five iterations each. Error bars represent one standard deviation, computed across participants. (C-D) Creativity and diversity gain: The improvement in measured creativity and diversity from the first iteration to the last.} 
\label{creativity} 
\end{figure*}

\subsection{Network Experiments on Collective Creativity}
We conducted large-scale online experiments on collective creativity, where participants (either human or AI) were embedded in a 5$\times$5 social network and engaged in a creative storytelling task. Participants selected and modified stories from their neighbours in the network, and their creations were presented to other participants over 25 iterations (Figure~\ref{over}A). The same prompt was used for both human and AI agents: \textit{Please creatively elaborate on the story, adding your own details and ideas}''. Participants were randomly assigned to one location (node) in the network and only contributed once to the experiment. All networks were initialized with the same story:
\begin{quote}
        ``\textit{As John reached for his front door, he realized his key was missing. Panic set in as he searched his pockets, but the key was nowhere to be found. Feeling defeated, he slumped against the door, only to hear a jingle from inside—his cat had been playing with the key all along.}''
\end{quote}

We compared three experimental conditions (Figure ~\ref{over}B): (1) \textit{human-only}, where all nodes were occupied by human participants (N=625), (2) \textit{AI-only}, where all nodes were simulated using GPT-4o (625 calls to the OpenAI API, and (3) \textit{human-AI}, consisting of an equal distribution of human participants (N=254, 50\%) and GPT-4o agents (50\%). To minimize bias, participants were not informed that any of the stories may have been generated by AI agents.

All experiments were conducted using PsyNet\footnote{\url{www.psynet.dev}}, a Python-based framework for advanced online psychological experiments~\cite{harrison2020gibbs}.

\subsection{Measures of Creativity}
To assess the quality of creations across conditions and iterations, we conducted a validation study with a separate group of 100 human participants. Each participant was presented with a randomized selection of stories from all experimental conditions and iterations and asked to rate their creativity on a 5-point scale, ranging from 1 (not creative at all) to 5 (extremely creative). Each participant evaluated 20 different stories. Crucially, participants were not informed whether stories were created by humans or AI. 

In addition to subjective creativity, we measured the diversity of stories by computing their semantic similarity. Specifically for the diversity analysis, we embedded each story using a TF-IDF vectorization approach \cite{leskovec2020mining} and calculated the pairwise cosine similarity between all stories within predefined iteration groups. Diversity was operationalized as the inverse of the mean cosine similarity, with lower similarity indicating greater diversity.

\section{Results}
\subsection{Creativity and Diversity}
We begin by examining the dynamics of collective creativity across the three experimental conditions. Figure~\ref{creativity}A shows the average creativity ratings from the validation study across the three conditions over time. The \textit{AI-only} condition exhibited the highest creativity rating ($M = 3.571,\, SD = 1.026$, 95\% CI: [3.491, 3.652]), significantly higher than both \textit{Human-only} and \textit{Human-AI} conditions ($p < .001$). The average creativity ratings in the \textit{Human-only} condition ($M = 2.482,\, SD = 1.026$, 95\% CI: [2.389, 2.576]) were similar to the \textit{Human-AI} condition ($M = 2.327,\, SD = 1.159$, 95\% CI: [2.184, 2.576]; $p < .001$). This result demostrates that GPT easily surpasses humans in simple creative writing tasks, even when evaluated by human raters. Interestingly, AI-only and Human-AI networks consistently improved over time, showing a significantly positive gain by the end of the experiment (GPT: $M = 0.464$, $SD = 0.154$, $p < 0.001$, Human-AI: $M = 0.264$, $SD = 0.175$, $p < 0.001$), but Human-only networks did not improve over time ($M = -0.133$, $SD = 0.067$, $p = 1.000$) (Figure \ref{creativity}C).

Next, we looked at the diversity of collective creations (Figure~\ref{creativity}B and D). Here, the \textit{AI-only} condition exhibited the highest diversity ($M = 0.880, SD = 0.017$, 95\% CI: [0.859, 0.900]), followed by \textit{Human-AI} ($M = 0.860, SD = 0.043$, 95\% CI: [0.806, 0.913]), and \textit{Human-only}, which showed the lowest diversity ($M = 0.823, SD = 0.019$, 95\% CI: [0.800, 0.900]). 

However, the evolution of collective creativity revealed an intriguing finding. Initially (iteration 1-5), stories in the \textit{AI-only} network were the highest in creativity and diversity, but over iterations, diversity steadily declined, with a drop of $M = -0.034$,\, $SD =0.17$, 95\% CI: [-0.047, -0.021] (Figure \ref{creativity}D). In contrast, the \textit{Human-AI} network started with lower collective diversity (similar levels to \textit{Human-only}) but exhibited the largest increase over time ($M = 0.098$, $SD =0.039$, 95\% CI: [0.064, 0.131]), ultimately achieving the highest overall diversity score in the final iterations.

\begin{figure}[ht]
\includegraphics[width=0.9\columnwidth]{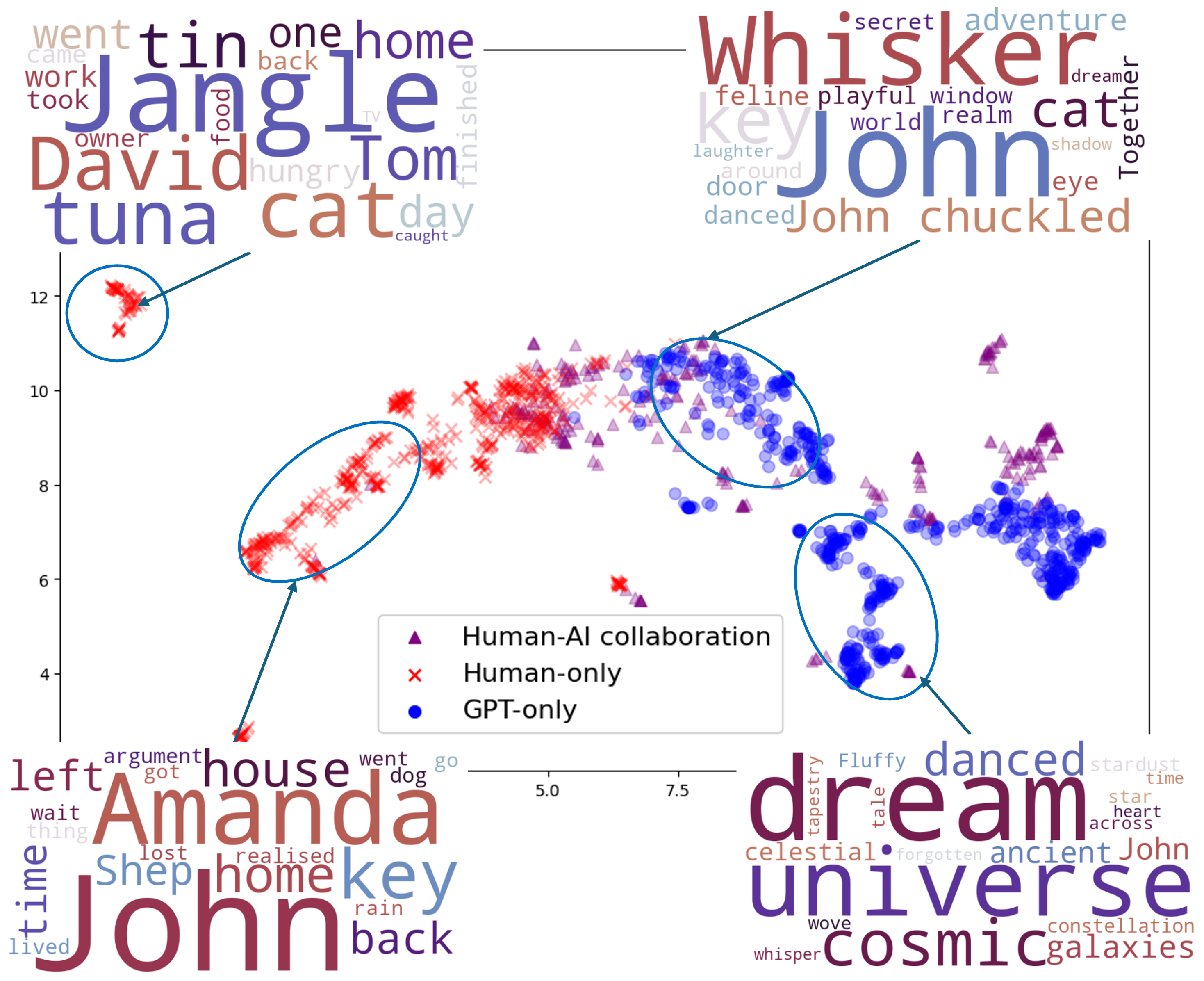}
\caption{UMAP projection of the shared semantic embedding space, highlighting word clouds for specific clusters.}
\label{umap}
\end{figure} 

\begin{figure*}[t]
\centering
\includegraphics[width=0.76\textwidth]{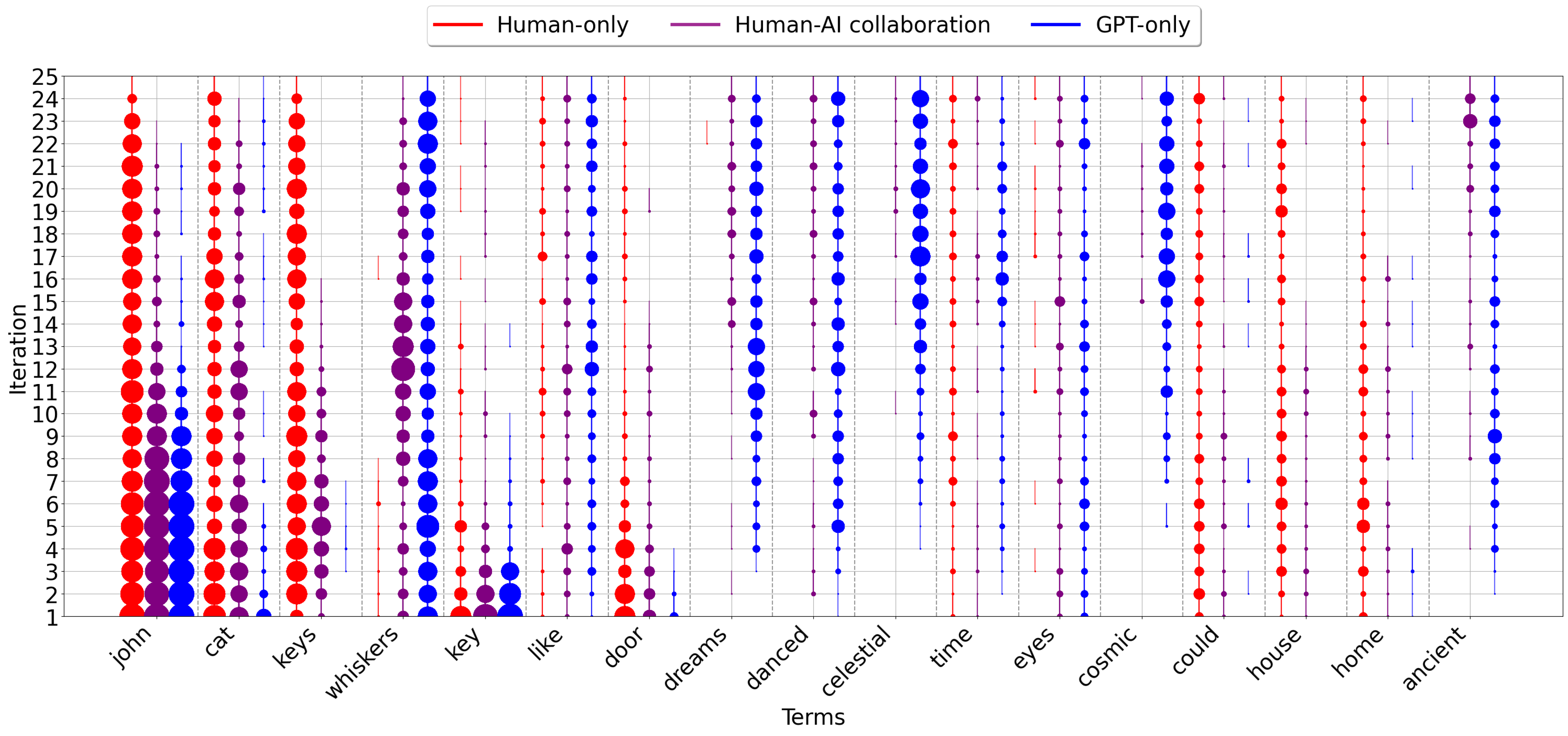}

\caption{Term dynamics by condition: Words are plotted along the horizontal axis, and generations are plotted along the vertical axis. A circle at position $(x, y)$ indicates that the word $x$ was used in a story at iteration $y$. A line denotes that the corresponding word was used by successive iterations. The size of the circle represents the frequency of the word’s appearance in the same iteration.}
\label{term}
\end{figure*}

\subsection{Semantic Analysis of Stories}

To visualize the difference between human and AI-generated stories, we created a semantic embedding space based on all written stories. Figure~\ref{umap} shows the UMAP embedding of the stories, visualizing the dynamics between the three conditions. Each point represents a single story, and the proximity of points reflects their semantic similarity in the original high-dimensional space. To compute the high-dimensional space used for the UMAP projection, we first generated semantic embeddings for each story using a pre-trained transformer model (all-MiniLM-L6-v2; \citeNP{wang2020minilm}). We then applied UMAP to reduce these embeddings to two dimensions, allowing us to visualize the semantic similarities among the stories.

Three key insights emerge from this analysis. First, the \textit{Human-only} and \textit{AI-only} conditions exhibit distinct clusters, indicating a marked difference in the language and semantics employed in their stories. The word cloud on the bottom right of Figure \ref{umap} shows that space-related words such as ``universe'' and ``cosmic'' are created when AI are involved in story generation. The word clouds on the left of Figure~\ref{umap} (largely composed from human data), on the other hand, suggest that story themes remain close to the original seed story (e.g., they share many of the same keywords, such as ``John'' and ``Key''). Second, the \textit{Human-only} condition exhibited additional human names such as “Amanda”, ``David'', and ``Tom'' suggesting potential shifts in the protagonist or narrative content, and indicating a distinct subgroup within the generated stories.
Third, in the \textit{Human-AI} condition, humans adopted a compromise in vocabulary and semantics, resulting in narratives that interpolate between the twe other conditions, suggesting that humans can be influenced by new ideas generated by AI. 
These clusters represent emergent ``niches'' in storytelling styles, characterized by unique interactions between human creativity and AI-driven semantic expansion.

Finally, to understand the shifts in narrative dynamics, we computed the frequency of dominant words in the story chracterized by TF-IDF scores. We combined all stories across conditions, calculated the TF-IDF for each term, and sorted them so that the most frequent terms are identified.
Figure~\ref{term} shows the dynamics of change in the narrative. Each  chain shows specific term prevalence over iterations, with circle size indicating the frequency of the term in the iteration. We can see that the longevity of specific keywords vary significantly depending on the condition. For example, certain words such as ``dreams'', ``danced'' and ``celestial'' only emerge and persist under the \textit{AI-only} condition. Conversely, words from the original story, including ``John'',  ``cat'', and ``keys'' persist until the end of the experiment only in the \textit{Human-only} condition. This suggests that humans tend to create new narratives that remain closely aligned with the original storyline, while AI outputs exhibited a unique tendency to converge on certain creative themes, such as space-related narratives, which were consistent across iterations. This convergence, while creative, can potentially indicate  difference in how AI agents interpert the task compared with humans.

\section{Discussion}
This study explored how the interplay between human and AI creativity in a storytelling task influences collective creative in experimental social networks. We examined three distinct network configurations -- human-only, AI-only, and human–AI -- to understand whether human-AI collaboration led people towards greater creativity. 

Our findings reveal that, from the outset, AI-only networks exhibited the highest levels of creativity and diversity compared to human-only and mixed networks. Over successive iterations, the creativity advantage of AI-only networks remained robust, suggesting that GPT’s ability to generate novel and divergent narratives was particularly well-suited to this task. However, while AI-generated stories initially displayed high diversity, this gradually declined over time. This suggests that although GPT can introduce novel ideas, it also exhibits a form of thematic convergence over time, leading to a reduction in overall diversity. This result complements previous evidence showing that while AI can enhance individual creativity, it may simultaneously reduce collective diversity \cite{doshi2024generative}.

In contrast, human-AI networks, which initially demonstrated lower diversity, ended up surpassing the diversity generated by AI-only networks. This shift was primarily driven by the distinct ways in which humans and AI approached the task. While AI frequently disregarded core narrative elements in favor of novel inventions, human participants tended to retain key story components, such as character identities (e.g., maintaining the protagonist, John) and objects (e.g., `keys'). The interplay between humans and AI in mixed networks enabled a dynamic balance between stability and novelty, enabling them to produce increasingly diverse outputs over time. This suggests that collaboration between humans and AI can harness the strengths of both, leading to richer and more diverse creative outcomes.

A crucial consideration in interpreting our results is the nature of the creative task itself. The storytelling task used in this study aligns with those traditionally employed in human creativity research \cite{kaufman2010cambridge, beaty2023associative}. However, the task proved to be relatively easy for a large language model like GPT, which consistently outperformed humans alone both in creativity and diversity. This does not necessarily imply that AI is intrinsically more creative than humans; rather, it highlights the limitations of using tasks designed for human cognition as benchmarks for AI creativity. AI’s advantage may stem from its ability to rapidly generate highly novel yet semantically coherent text, a skill optimized by its training on vast datasets \cite{burton2024large, brinkmann2023machine}. Conversely, human creativity and creative writing is influenced by memory constraints, cognitive biases, and adherence to implicit narrative conventions \cite{hayes2012modeling}. 

Moving forward, creativity research should move beyond simple tasks that may be well-suited for assessing individual human creativity but fall short in evaluating AI capabilities or capturing the dynamics of human–AI synergy. For example, using multimodal creative tasks that do not only rely on language, such as music creation \cite{anglada2023large} and drawings \cite{kumar2024comparing, hart2017creative, hart2018creative}. Moreover, future studies should focus on designing complex tasks that challenge both humans and AI in meaningful ways, allowing for a more nuanced understanding of collective creativity and underlying mechanisms.

\section{Limitations and conclusion}
Our study aimed to simulate social-naturalistic dynamics within experimental societies. However, our approach diverges from how ecologically valid social interactions unfold in the real world. Unlike our experiments, where participants engaged only once, individuals in real-world settings continuously interact with each other over time. Additionally, real-world exchanges are far more complex than the constrained task of creating simple stories used in our experiment. Another key limitation is the static structure of our simulated social network, which used a grid-based topology. Real social networks, such as groups of friends or work relationships, tend to have a modular structure with irregular patterns and evolving connections, allowing individuals to actively shape their interactions by forming and adjusting their social ties rather than being confined to static connections. Finally, our paradigm avoids any direct communication between participants and masks agent information, such as whether your partner is a human or an AI. Prior research suggests that awareness of interacting with an AI system may alter individual behavior, which in turn could influence collective dynamics \cite{tsvetkova2024new}.

In addition, our study has several technical limitations. First, key parameters of the AI models—such as model architecture, temperature settings, and prompt design—may have influenced the results. We intend to systematically examine the impact of these variables in future work. Second, while we employed large language models (LLMs) for story generation and selection, more advanced AI agents now enable richer forms of interaction with both the environment and other agents—an aspect not addressed in the current study \cite{google2025a2a}. Lastly, our validation approach was relatively limited. We relied on a single measure of creativity, and participants were not provided with an explicit definition of the term. Future studies could incorporate a more comprehensive set of creativity metrics and a more robust validation framework.

Despite these limitations, none of these constraints are inherent to our methodological approach. Our framework can be extended to incorporate more ecologically valid dynamics, including persistent interactions, evolving social structures, and greater participant agency. Our approach can also be extended to non-verbal, open-ended creative task, such as music and visual art. While our study represents only an initial exploration of hybrid experimental social networks, it highlights the vast potential of this approach for future research in cognitive and computer science.

\bibliographystyle{apacite}

\setlength{\bibleftmargin}{.125in}
\setlength{\bibindent}{-\bibleftmargin}

\bibliography{CogSci_Template}

\begin{thebibliography}{}

\bibitem [\protect \citeauthoryear {%
Amabile%
}{%
Amabile%
}{%
{\protect \APACyear {1982}}%
}]{%
amabile1982social}
\APACinsertmetastar {%
amabile1982social}%
\begin{APACrefauthors}%
Amabile, T\BPBI M.%
\end{APACrefauthors}%
\unskip\
\newblock
\APACrefYearMonthDay{1982}{}{}.
\newblock
{\BBOQ}\APACrefatitle {Social psychology of creativity: A consensual assessment technique.} {Social psychology of creativity: A consensual assessment technique.}{\BBCQ}
\newblock
\APACjournalVolNumPages{Journal of personality and social psychology}{43}{5}{997}.
\PrintBackRefs{\CurrentBib}

\bibitem [\protect \citeauthoryear {%
Anderson%
, Shah%
\BCBL {}\ \BBA {} Kreminski%
}{%
Anderson%
\ \protect \BOthers {.}}{%
{\protect \APACyear {2024}}%
}]{%
anderson2024homogenization}
\APACinsertmetastar {%
anderson2024homogenization}%
\begin{APACrefauthors}%
Anderson, B\BPBI R.%
, Shah, J\BPBI H.%
\BCBL {}\ \BBA {} Kreminski, M.%
\end{APACrefauthors}%
\unskip\
\newblock
\APACrefYearMonthDay{2024}{}{}.
\newblock
{\BBOQ}\APACrefatitle {Homogenization effects of large language models on human creative ideation} {Homogenization effects of large language models on human creative ideation}.{\BBCQ}
\newblock
\BIn{} \APACrefbtitle {Proceedings of the 16th conference on creativity \& cognition} {Proceedings of the 16th conference on creativity \& cognition}\ (\BPGS\ 413--425).
\PrintBackRefs{\CurrentBib}

\bibitem [\protect \citeauthoryear {%
Anglada-Tort%
, Harrison%
, Lee%
\BCBL {}\ \BBA {} Jacoby%
}{%
Anglada-Tort%
\ \protect \BOthers {.}}{%
{\protect \APACyear {2023}}%
}]{%
anglada2023large}
\APACinsertmetastar {%
anglada2023large}%
\begin{APACrefauthors}%
Anglada-Tort, M.%
, Harrison, P\BPBI M.%
, Lee, H.%
\BCBL {}\ \BBA {} Jacoby, N.%
\end{APACrefauthors}%
\unskip\
\newblock
\APACrefYearMonthDay{2023}{}{}.
\newblock
{\BBOQ}\APACrefatitle {Large-scale iterated singing experiments reveal oral transmission mechanisms underlying music evolution} {Large-scale iterated singing experiments reveal oral transmission mechanisms underlying music evolution}.{\BBCQ}
\newblock
\APACjournalVolNumPages{Current Biology}{33}{8}{1472--1486}.
\PrintBackRefs{\CurrentBib}

\bibitem [\protect \citeauthoryear {%
Bansal%
\ \protect \BOthers {.}}{%
Bansal%
\ \protect \BOthers {.}}{%
{\protect \APACyear {2021}}%
}]{%
bansal2021does}
\APACinsertmetastar {%
bansal2021does}%
\begin{APACrefauthors}%
Bansal, G.%
, Wu, T.%
, Zhou, J.%
, Fok, R.%
, Nushi, B.%
, Kamar, E.%
\BDBL {}Weld, D.%
\end{APACrefauthors}%
\unskip\
\newblock
\APACrefYearMonthDay{2021}{}{}.
\newblock
{\BBOQ}\APACrefatitle {Does the whole exceed its parts? the effect of ai explanations on complementary team performance} {Does the whole exceed its parts? the effect of ai explanations on complementary team performance}.{\BBCQ}
\newblock
\BIn{} \APACrefbtitle {Proceedings of the 2021 CHI conference on human factors in computing systems} {Proceedings of the 2021 chi conference on human factors in computing systems}\ (\BPGS\ 1--16).
\PrintBackRefs{\CurrentBib}

\bibitem [\protect \citeauthoryear {%
Beaty%
\ \BBA {} Kenett%
}{%
Beaty%
\ \BBA {} Kenett%
}{%
{\protect \APACyear {2023}}%
}]{%
beaty2023associative}
\APACinsertmetastar {%
beaty2023associative}%
\begin{APACrefauthors}%
Beaty, R\BPBI E.%
\BCBT {}\ \BBA {} Kenett, Y\BPBI N.%
\end{APACrefauthors}%
\unskip\
\newblock
\APACrefYearMonthDay{2023}{}{}.
\newblock
{\BBOQ}\APACrefatitle {Associative thinking at the core of creativity} {Associative thinking at the core of creativity}.{\BBCQ}
\newblock
\APACjournalVolNumPages{Trends in cognitive sciences}{27}{7}{671--683}.
\PrintBackRefs{\CurrentBib}

\bibitem [\protect \citeauthoryear {%
Bengio%
\ \protect \BOthers {.}}{%
Bengio%
\ \protect \BOthers {.}}{%
{\protect \APACyear {2024}}%
}]{%
bengio2024managing}
\APACinsertmetastar {%
bengio2024managing}%
\begin{APACrefauthors}%
Bengio, Y.%
, Hinton, G.%
, Yao, A.%
, Song, D.%
, Abbeel, P.%
, Darrell, T.%
\BDBL {}others%
\end{APACrefauthors}%
\unskip\
\newblock
\APACrefYearMonthDay{2024}{}{}.
\newblock
{\BBOQ}\APACrefatitle {Managing extreme AI risks amid rapid progress} {Managing extreme ai risks amid rapid progress}.{\BBCQ}
\newblock
\APACjournalVolNumPages{Science}{384}{6698}{842--845}.
\PrintBackRefs{\CurrentBib}

\bibitem [\protect \citeauthoryear {%
Boden%
}{%
Boden%
}{%
{\protect \APACyear {2004}}%
}]{%
boden2004creative}
\APACinsertmetastar {%
boden2004creative}%
\begin{APACrefauthors}%
Boden, M\BPBI A.%
\end{APACrefauthors}%
\unskip\
\newblock
\APACrefYear{2004}.
\newblock
\APACrefbtitle {The creative mind: Myths and mechanisms} {The creative mind: Myths and mechanisms}.
\newblock
\APACaddressPublisher{}{Routledge}.
\PrintBackRefs{\CurrentBib}

\bibitem [\protect \citeauthoryear {%
Brinkmann%
\ \protect \BOthers {.}}{%
Brinkmann%
\ \protect \BOthers {.}}{%
{\protect \APACyear {2023}}%
}]{%
brinkmann2023machine}
\APACinsertmetastar {%
brinkmann2023machine}%
\begin{APACrefauthors}%
Brinkmann, L.%
, Baumann, F.%
, Bonnefon, J\BHBI F.%
, Derex, M.%
, M{\"u}ller, T\BPBI F.%
, Nussberger, A\BHBI M.%
\BDBL {}others%
\end{APACrefauthors}%
\unskip\
\newblock
\APACrefYearMonthDay{2023}{}{}.
\newblock
{\BBOQ}\APACrefatitle {Machine culture} {Machine culture}.{\BBCQ}
\newblock
\APACjournalVolNumPages{Nature Human Behaviour}{7}{11}{1855--1868}.
\PrintBackRefs{\CurrentBib}

\bibitem [\protect \citeauthoryear {%
Brown%
\ \protect \BOthers {.}}{%
Brown%
\ \protect \BOthers {.}}{%
{\protect \APACyear {2020}}%
}]{%
brown2020language}
\APACinsertmetastar {%
brown2020language}%
\begin{APACrefauthors}%
Brown, T.%
, Mann, B.%
, Ryder, N.%
, Subbiah, M.%
, Kaplan, J\BPBI D.%
, Dhariwal, P.%
\BDBL {}others%
\end{APACrefauthors}%
\unskip\
\newblock
\APACrefYearMonthDay{2020}{}{}.
\newblock
{\BBOQ}\APACrefatitle {Language models are few-shot learners} {Language models are few-shot learners}.{\BBCQ}
\newblock
\APACjournalVolNumPages{Advances in neural information processing systems}{33}{}{1877--1901}.
\PrintBackRefs{\CurrentBib}

\bibitem [\protect \citeauthoryear {%
Burton%
\ \protect \BOthers {.}}{%
Burton%
\ \protect \BOthers {.}}{%
{\protect \APACyear {2024}}%
}]{%
burton2024large}
\APACinsertmetastar {%
burton2024large}%
\begin{APACrefauthors}%
Burton, J\BPBI W.%
, Lopez-Lopez, E.%
, Hechtlinger, S.%
, Rahwan, Z.%
, Aeschbach, S.%
, Bakker, M\BPBI A.%
\BDBL {}others%
\end{APACrefauthors}%
\unskip\
\newblock
\APACrefYearMonthDay{2024}{}{}.
\newblock
{\BBOQ}\APACrefatitle {How large language models can reshape collective intelligence} {How large language models can reshape collective intelligence}.{\BBCQ}
\newblock
\APACjournalVolNumPages{Nature Human Behaviour}{8}{9}{1643--1655}.
\PrintBackRefs{\CurrentBib}

\bibitem [\protect \citeauthoryear {%
Centola%
}{%
Centola%
}{%
{\protect \APACyear {2022}}%
}]{%
centola2022network}
\APACinsertmetastar {%
centola2022network}%
\begin{APACrefauthors}%
Centola, D.%
\end{APACrefauthors}%
\unskip\
\newblock
\APACrefYearMonthDay{2022}{}{}.
\newblock
{\BBOQ}\APACrefatitle {The network science of collective intelligence} {The network science of collective intelligence}.{\BBCQ}
\newblock
\APACjournalVolNumPages{Trends in Cognitive Sciences}{26}{11}{923--941}.
\PrintBackRefs{\CurrentBib}

\bibitem [\protect \citeauthoryear {%
Centola%
\ \BBA {} Baronchelli%
}{%
Centola%
\ \BBA {} Baronchelli%
}{%
{\protect \APACyear {2015}}%
}]{%
centola2015spontaneous}
\APACinsertmetastar {%
centola2015spontaneous}%
\begin{APACrefauthors}%
Centola, D.%
\BCBT {}\ \BBA {} Baronchelli, A.%
\end{APACrefauthors}%
\unskip\
\newblock
\APACrefYearMonthDay{2015}{}{}.
\newblock
{\BBOQ}\APACrefatitle {The spontaneous emergence of conventions: An experimental study of cultural evolution} {The spontaneous emergence of conventions: An experimental study of cultural evolution}.{\BBCQ}
\newblock
\APACjournalVolNumPages{Proceedings of the National Academy of Sciences}{112}{7}{1989--1994}.
\PrintBackRefs{\CurrentBib}

\bibitem [\protect \citeauthoryear {%
Chen%
, Liao%
, Wortman~Vaughan%
\BCBL {}\ \BBA {} Bansal%
}{%
Chen%
\ \protect \BOthers {.}}{%
{\protect \APACyear {2023}}%
}]{%
chen2023understanding}
\APACinsertmetastar {%
chen2023understanding}%
\begin{APACrefauthors}%
Chen, V.%
, Liao, Q\BPBI V.%
, Wortman~Vaughan, J.%
\BCBL {}\ \BBA {} Bansal, G.%
\end{APACrefauthors}%
\unskip\
\newblock
\APACrefYearMonthDay{2023}{}{}.
\newblock
{\BBOQ}\APACrefatitle {Understanding the role of human intuition on reliance in human-AI decision-making with explanations} {Understanding the role of human intuition on reliance in human-ai decision-making with explanations}.{\BBCQ}
\newblock
\APACjournalVolNumPages{Proceedings of the ACM on Human-computer Interaction}{7}{CSCW2}{1--32}.
\PrintBackRefs{\CurrentBib}

\bibitem [\protect \citeauthoryear {%
Collins%
\ \protect \BOthers {.}}{%
Collins%
\ \protect \BOthers {.}}{%
{\protect \APACyear {2024}}%
}]{%
collins2024building}
\APACinsertmetastar {%
collins2024building}%
\begin{APACrefauthors}%
Collins, K\BPBI M.%
, Sucholutsky, I.%
, Bhatt, U.%
, Chandra, K.%
, Wong, L.%
, Lee, M.%
\BDBL {}others%
\end{APACrefauthors}%
\unskip\
\newblock
\APACrefYearMonthDay{2024}{}{}.
\newblock
{\BBOQ}\APACrefatitle {Building machines that learn and think with people} {Building machines that learn and think with people}.{\BBCQ}
\newblock
\APACjournalVolNumPages{Nature Human Behaviour}{8}{10}{1851--1863}.
\PrintBackRefs{\CurrentBib}

\bibitem [\protect \citeauthoryear {%
Colton%
\ \BBA {} Wiggins%
}{%
Colton%
\ \BBA {} Wiggins%
}{%
{\protect \APACyear {2012}}%
}]{%
colton2012computational}
\APACinsertmetastar {%
colton2012computational}%
\begin{APACrefauthors}%
Colton, S.%
\BCBT {}\ \BBA {} Wiggins, G\BPBI A.%
\end{APACrefauthors}%
\unskip\
\newblock
\APACrefYearMonthDay{2012}{}{}.
\newblock
{\BBOQ}\APACrefatitle {Computational creativity: The final frontier?} {Computational creativity: The final frontier?}{\BBCQ}
\newblock
\BIn{} \APACrefbtitle {ECAI 2012} {Ecai 2012}\ (\BPGS\ 21--26).
\newblock
\APACaddressPublisher{}{IOS Press}.
\PrintBackRefs{\CurrentBib}

\bibitem [\protect \citeauthoryear {%
Derex%
, Beugin%
, Godelle%
\BCBL {}\ \BBA {} Raymond%
}{%
Derex%
\ \protect \BOthers {.}}{%
{\protect \APACyear {2013}}%
}]{%
derex2013experimental}
\APACinsertmetastar {%
derex2013experimental}%
\begin{APACrefauthors}%
Derex, M.%
, Beugin, M\BHBI P.%
, Godelle, B.%
\BCBL {}\ \BBA {} Raymond, M.%
\end{APACrefauthors}%
\unskip\
\newblock
\APACrefYearMonthDay{2013}{}{}.
\newblock
{\BBOQ}\APACrefatitle {Experimental evidence for the influence of group size on cultural complexity} {Experimental evidence for the influence of group size on cultural complexity}.{\BBCQ}
\newblock
\APACjournalVolNumPages{Nature}{503}{7476}{389--391}.
\PrintBackRefs{\CurrentBib}

\bibitem [\protect \citeauthoryear {%
Doshi%
\ \BBA {} Hauser%
}{%
Doshi%
\ \BBA {} Hauser%
}{%
{\protect \APACyear {2024}}%
}]{%
doshi2024generative}
\APACinsertmetastar {%
doshi2024generative}%
\begin{APACrefauthors}%
Doshi, A\BPBI R.%
\BCBT {}\ \BBA {} Hauser, O\BPBI P.%
\end{APACrefauthors}%
\unskip\
\newblock
\APACrefYearMonthDay{2024}{}{}.
\newblock
{\BBOQ}\APACrefatitle {Generative AI enhances individual creativity but reduces the collective diversity of novel content} {Generative ai enhances individual creativity but reduces the collective diversity of novel content}.{\BBCQ}
\newblock
\APACjournalVolNumPages{Science Advances}{10}{28}{eadn5290}.
\PrintBackRefs{\CurrentBib}

\bibitem [\protect \citeauthoryear {%
Epstein%
\ \protect \BOthers {.}}{%
Epstein%
\ \protect \BOthers {.}}{%
{\protect \APACyear {2023}}%
}]{%
epstein2023art}
\APACinsertmetastar {%
epstein2023art}%
\begin{APACrefauthors}%
Epstein, Z.%
, Hertzmann, A.%
, of Human~Creativity, I.%
, Akten, M.%
, Farid, H.%
, Fjeld, J.%
\BDBL {}others%
\end{APACrefauthors}%
\unskip\
\newblock
\APACrefYearMonthDay{2023}{}{}.
\newblock
{\BBOQ}\APACrefatitle {Art and the science of generative AI} {Art and the science of generative ai}.{\BBCQ}
\newblock
\APACjournalVolNumPages{Science}{380}{6650}{1110--1111}.
\PrintBackRefs{\CurrentBib}

\bibitem [\protect \citeauthoryear {%
{Google Developers Blog}%
}{%
{Google Developers Blog}%
}{%
{\protect \APACyear {2025}}%
}]{%
google2025a2a}
\APACinsertmetastar {%
google2025a2a}%
\begin{APACrefauthors}%
{Google Developers Blog}.%
\end{APACrefauthors}%
\unskip\
\newblock
\APACrefYearMonthDay{2025}{}{}.
\newblock
\APACrefbtitle {Announcing the agent2agent protocol (a2a).} {Announcing the agent2agent protocol (a2a).}
\PrintBackRefs{\CurrentBib}

\bibitem [\protect \citeauthoryear {%
Guilford%
}{%
Guilford%
}{%
{\protect \APACyear {1967}}%
}]{%
guilford1967nature}
\APACinsertmetastar {%
guilford1967nature}%
\begin{APACrefauthors}%
Guilford, J\BPBI P.%
\end{APACrefauthors}%
\unskip\
\newblock
\APACrefYearMonthDay{1967}{}{}.
\newblock
{\BBOQ}\APACrefatitle {The nature of human intelligence} {The nature of human intelligence}.{\BBCQ}
\newblock
\APACjournalVolNumPages{New York: Macgraw Hill}{}{}{}.
\PrintBackRefs{\CurrentBib}

\bibitem [\protect \citeauthoryear {%
Harrison%
\ \protect \BOthers {.}}{%
Harrison%
\ \protect \BOthers {.}}{%
{\protect \APACyear {2020}}%
}]{%
harrison2020gibbs}
\APACinsertmetastar {%
harrison2020gibbs}%
\begin{APACrefauthors}%
Harrison, P.%
, Marjieh, R.%
, Adolfi, F.%
, van Rijn, P.%
, Anglada-Tort, M.%
, Tchernichovski, O.%
\BDBL {}Jacoby, N.%
\end{APACrefauthors}%
\unskip\
\newblock
\APACrefYearMonthDay{2020}{}{}.
\newblock
{\BBOQ}\APACrefatitle {Gibbs sampling with people} {Gibbs sampling with people}.{\BBCQ}
\newblock
\APACjournalVolNumPages{Advances in neural information processing systems}{33}{}{10659--10671}.
\PrintBackRefs{\CurrentBib}

\bibitem [\protect \citeauthoryear {%
Hart%
\ \protect \BOthers {.}}{%
Hart%
\ \protect \BOthers {.}}{%
{\protect \APACyear {2018}}%
}]{%
hart2018creative}
\APACinsertmetastar {%
hart2018creative}%
\begin{APACrefauthors}%
Hart, Y.%
, Goldberg, H.%
, Striem-Amit, E.%
, Mayo, A\BPBI E.%
, Noy, L.%
\BCBL {}\ \BBA {} Alon, U.%
\end{APACrefauthors}%
\unskip\
\newblock
\APACrefYearMonthDay{2018}{}{}.
\newblock
{\BBOQ}\APACrefatitle {Creative exploration as a scale-invariant search on a meaning landscape} {Creative exploration as a scale-invariant search on a meaning landscape}.{\BBCQ}
\newblock
\APACjournalVolNumPages{Nature Communications}{9}{1}{5411}.
\PrintBackRefs{\CurrentBib}

\bibitem [\protect \citeauthoryear {%
Hart%
\ \protect \BOthers {.}}{%
Hart%
\ \protect \BOthers {.}}{%
{\protect \APACyear {2017}}%
}]{%
hart2017creative}
\APACinsertmetastar {%
hart2017creative}%
\begin{APACrefauthors}%
Hart, Y.%
, Mayo, A\BPBI E.%
, Mayo, R.%
, Rozenkrantz, L.%
, Tendler, A.%
, Alon, U.%
\BCBL {}\ \BBA {} Noy, L.%
\end{APACrefauthors}%
\unskip\
\newblock
\APACrefYearMonthDay{2017}{}{}.
\newblock
{\BBOQ}\APACrefatitle {Creative foraging: An experimental paradigm for studying exploration and discovery} {Creative foraging: An experimental paradigm for studying exploration and discovery}.{\BBCQ}
\newblock
\APACjournalVolNumPages{PloS one}{12}{8}{e0182133}.
\PrintBackRefs{\CurrentBib}

\bibitem [\protect \citeauthoryear {%
Hayes%
}{%
Hayes%
}{%
{\protect \APACyear {2012}}%
}]{%
hayes2012modeling}
\APACinsertmetastar {%
hayes2012modeling}%
\begin{APACrefauthors}%
Hayes, J\BPBI R.%
\end{APACrefauthors}%
\unskip\
\newblock
\APACrefYearMonthDay{2012}{}{}.
\newblock
{\BBOQ}\APACrefatitle {Modeling and remodeling writing} {Modeling and remodeling writing}.{\BBCQ}
\newblock
\APACjournalVolNumPages{Written Communication}{29}{3}{369--388}.
\PrintBackRefs{\CurrentBib}

\bibitem [\protect \citeauthoryear {%
Hurst%
\ \protect \BOthers {.}}{%
Hurst%
\ \protect \BOthers {.}}{%
{\protect \APACyear {2024}}%
}]{%
hurst2024gpt}
\APACinsertmetastar {%
hurst2024gpt}%
\begin{APACrefauthors}%
Hurst, A.%
, Lerer, A.%
, Goucher, A\BPBI P.%
, Perelman, A.%
, Ramesh, A.%
, Clark, A.%
\BDBL {}others%
\end{APACrefauthors}%
\unskip\
\newblock
\APACrefYearMonthDay{2024}{}{}.
\newblock
{\BBOQ}\APACrefatitle {{GPT}-4o system card} {{GPT}-4o system card}.{\BBCQ}
\newblock
\APACjournalVolNumPages{arXiv preprint arXiv:2410.21276}{}{}{}.
\PrintBackRefs{\CurrentBib}

\bibitem [\protect \citeauthoryear {%
Kaufman%
\ \BBA {} Sternberg%
}{%
Kaufman%
\ \BBA {} Sternberg%
}{%
{\protect \APACyear {2010}}%
}]{%
kaufman2010cambridge}
\APACinsertmetastar {%
kaufman2010cambridge}%
\begin{APACrefauthors}%
Kaufman, J\BPBI C.%
\BCBT {}\ \BBA {} Sternberg, R\BPBI J.%
\end{APACrefauthors}%
\unskip\
\newblock
\APACrefYear{2010}.
\newblock
\APACrefbtitle {The Cambridge handbook of creativity} {The cambridge handbook of creativity}.
\newblock
\APACaddressPublisher{}{Cambridge University Press}.
\PrintBackRefs{\CurrentBib}

\bibitem [\protect \citeauthoryear {%
Kumar%
\ \protect \BOthers {.}}{%
Kumar%
\ \protect \BOthers {.}}{%
{\protect \APACyear {2024}}%
}]{%
kumar2024comparing}
\APACinsertmetastar {%
kumar2024comparing}%
\begin{APACrefauthors}%
Kumar, S.%
, Marjieh, R.%
, Zhang, B.%
, Campbell, D.%
, Hu, M\BPBI Y.%
, Bhatt, U.%
\BDBL {}Griffiths, T.%
\end{APACrefauthors}%
\unskip\
\newblock
\APACrefYearMonthDay{2024}{}{}.
\newblock
{\BBOQ}\APACrefatitle {Comparing Abstraction in Humans and Machines Using Multimodal Serial Reproduction} {Comparing abstraction in humans and machines using multimodal serial reproduction}.{\BBCQ}
\newblock
\BIn{} \APACrefbtitle {Proceedings of the Annual Meeting of the Cognitive Science Society} {Proceedings of the annual meeting of the cognitive science society}\ (\BVOL~46).
\PrintBackRefs{\CurrentBib}

\bibitem [\protect \citeauthoryear {%
Lamb%
, Brown%
\BCBL {}\ \BBA {} Clarke%
}{%
Lamb%
\ \protect \BOthers {.}}{%
{\protect \APACyear {2018}}%
}]{%
lamb2018evaluating}
\APACinsertmetastar {%
lamb2018evaluating}%
\begin{APACrefauthors}%
Lamb, C.%
, Brown, D\BPBI G.%
\BCBL {}\ \BBA {} Clarke, C\BPBI L.%
\end{APACrefauthors}%
\unskip\
\newblock
\APACrefYearMonthDay{2018}{}{}.
\newblock
{\BBOQ}\APACrefatitle {Evaluating computational creativity: An interdisciplinary tutorial} {Evaluating computational creativity: An interdisciplinary tutorial}.{\BBCQ}
\newblock
\APACjournalVolNumPages{ACM Computing Surveys (CSUR)}{51}{2}{1--34}.
\PrintBackRefs{\CurrentBib}

\bibitem [\protect \citeauthoryear {%
Lee%
\ \BBA {} Chung%
}{%
Lee%
\ \BBA {} Chung%
}{%
{\protect \APACyear {2024}}%
}]{%
lee2024empirical}
\APACinsertmetastar {%
lee2024empirical}%
\begin{APACrefauthors}%
Lee, B\BPBI C.%
\BCBT {}\ \BBA {} Chung, J.%
\end{APACrefauthors}%
\unskip\
\newblock
\APACrefYearMonthDay{2024}{}{}.
\newblock
{\BBOQ}\APACrefatitle {An empirical investigation of the impact of ChatGPT on creativity} {An empirical investigation of the impact of chatgpt on creativity}.{\BBCQ}
\newblock
\APACjournalVolNumPages{Nature Human Behaviour}{8}{10}{1906--1914}.
\PrintBackRefs{\CurrentBib}

\bibitem [\protect \citeauthoryear {%
Leskovec%
, Rajaraman%
\BCBL {}\ \BBA {} Ullman%
}{%
Leskovec%
\ \protect \BOthers {.}}{%
{\protect \APACyear {2020}}%
}]{%
leskovec2020mining}
\APACinsertmetastar {%
leskovec2020mining}%
\begin{APACrefauthors}%
Leskovec, J.%
, Rajaraman, A.%
\BCBL {}\ \BBA {} Ullman, J\BPBI D.%
\end{APACrefauthors}%
\unskip\
\newblock
\APACrefYear{2020}.
\newblock
\APACrefbtitle {Mining of massive data sets} {Mining of massive data sets}.
\newblock
\APACaddressPublisher{}{Cambridge university press}.
\PrintBackRefs{\CurrentBib}

\bibitem [\protect \citeauthoryear {%
Liu%
, Lai%
\BCBL {}\ \BBA {} Tan%
}{%
Liu%
\ \protect \BOthers {.}}{%
{\protect \APACyear {2021}}%
}]{%
liu2021understanding}
\APACinsertmetastar {%
liu2021understanding}%
\begin{APACrefauthors}%
Liu, H.%
, Lai, V.%
\BCBL {}\ \BBA {} Tan, C.%
\end{APACrefauthors}%
\unskip\
\newblock
\APACrefYearMonthDay{2021}{}{}.
\newblock
{\BBOQ}\APACrefatitle {Understanding the effect of out-of-distribution examples and interactive explanations on human-ai decision making} {Understanding the effect of out-of-distribution examples and interactive explanations on human-ai decision making}.{\BBCQ}
\newblock
\APACjournalVolNumPages{Proceedings of the ACM on Human-Computer Interaction}{5}{CSCW2}{1--45}.
\PrintBackRefs{\CurrentBib}

\bibitem [\protect \citeauthoryear {%
Maher%
}{%
Maher%
}{%
{\protect \APACyear {2012}}%
}]{%
maher2012computational}
\APACinsertmetastar {%
maher2012computational}%
\begin{APACrefauthors}%
Maher, M\BPBI L.%
\end{APACrefauthors}%
\unskip\
\newblock
\APACrefYearMonthDay{2012}{}{}.
\newblock
{\BBOQ}\APACrefatitle {Computational and collective creativity: Who's being creative?} {Computational and collective creativity: Who's being creative?}{\BBCQ}
\newblock
\BIn{} \APACrefbtitle {ICCC} {Iccc}\ (\BPGS\ 67--71).
\PrintBackRefs{\CurrentBib}

\bibitem [\protect \citeauthoryear {%
Malone%
\ \BBA {} Bernstein%
}{%
Malone%
\ \BBA {} Bernstein%
}{%
{\protect \APACyear {2022}}%
}]{%
malone2022handbook}
\APACinsertmetastar {%
malone2022handbook}%
\begin{APACrefauthors}%
Malone, T\BPBI W.%
\BCBT {}\ \BBA {} Bernstein, M.%
\end{APACrefauthors}%
\unskip\
\newblock
\APACrefYear{2022}.
\newblock
\APACrefbtitle {Handbook of collective intelligence} {Handbook of collective intelligence}.
\newblock
\APACaddressPublisher{}{MIT Press}.
\PrintBackRefs{\CurrentBib}

\bibitem [\protect \citeauthoryear {%
Marjieh%
, Anglada-Tort%
, Griffiths%
\BCBL {}\ \BBA {} Jacoby%
}{%
Marjieh%
\ \protect \BOthers {.}}{%
{\protect \APACyear {2025}}%
}]{%
marjieh2025characterizing}
\APACinsertmetastar {%
marjieh2025characterizing}%
\begin{APACrefauthors}%
Marjieh, R.%
, Anglada-Tort, M.%
, Griffiths, T\BPBI L.%
\BCBL {}\ \BBA {} Jacoby, N.%
\end{APACrefauthors}%
\unskip\
\newblock
\APACrefYearMonthDay{2025}{}{}.
\newblock
{\BBOQ}\APACrefatitle {Characterizing the Interaction of Cultural Evolution Mechanisms in Experimental Social Networks} {Characterizing the interaction of cultural evolution mechanisms in experimental social networks}.{\BBCQ}
\newblock
\APACjournalVolNumPages{arXiv preprint arXiv:2502.12847}{}{}{}.
\PrintBackRefs{\CurrentBib}

\bibitem [\protect \citeauthoryear {%
Muller%
\ \protect \BOthers {.}}{%
Muller%
\ \protect \BOthers {.}}{%
{\protect \APACyear {2024}}%
}]{%
muller2024group}
\APACinsertmetastar {%
muller2024group}%
\begin{APACrefauthors}%
Muller, M.%
, Houde, S.%
, Gonzalez, G.%
, Brimijoin, K.%
, Ross, S\BPBI I.%
, Moran, D\BPBI A\BPBI S.%
\BCBL {}\ \BBA {} Weisz, J\BPBI D.%
\end{APACrefauthors}%
\unskip\
\newblock
\APACrefYearMonthDay{2024}{}{}.
\newblock
{\BBOQ}\APACrefatitle {Group Brainstorming with an AI Agent: Creating and Selecting Ideas} {Group brainstorming with an ai agent: Creating and selecting ideas}.{\BBCQ}
\newblock
\BIn{} \APACrefbtitle {International Conference on Computational Creativity.} {International conference on computational creativity.}
\PrintBackRefs{\CurrentBib}

\bibitem [\protect \citeauthoryear {%
Parjanen%
}{%
Parjanen%
}{%
{\protect \APACyear {2012}}%
}]{%
parjanen2012experiencing}
\APACinsertmetastar {%
parjanen2012experiencing}%
\begin{APACrefauthors}%
Parjanen, S.%
\end{APACrefauthors}%
\unskip\
\newblock
\APACrefYearMonthDay{2012}{}{}.
\newblock
{\BBOQ}\APACrefatitle {Experiencing creativity in the organization: From individual creativity to collective creativity.} {Experiencing creativity in the organization: From individual creativity to collective creativity.}{\BBCQ}
\newblock
\APACjournalVolNumPages{Interdisciplinary Journal of Information, Knowledge \& Management}{7}{}{}.
\PrintBackRefs{\CurrentBib}

\bibitem [\protect \citeauthoryear {%
Ramesh%
, Dhariwal%
, Nichol%
, Chu%
\BCBL {}\ \BBA {} Chen%
}{%
Ramesh%
\ \protect \BOthers {.}}{%
{\protect \APACyear {2022}}%
}]{%
ramesh2022hierarchical}
\APACinsertmetastar {%
ramesh2022hierarchical}%
\begin{APACrefauthors}%
Ramesh, A.%
, Dhariwal, P.%
, Nichol, A.%
, Chu, C.%
\BCBL {}\ \BBA {} Chen, M.%
\end{APACrefauthors}%
\unskip\
\newblock
\APACrefYearMonthDay{2022}{}{}.
\newblock
{\BBOQ}\APACrefatitle {Hierarchical text-conditional image generation with clip latents} {Hierarchical text-conditional image generation with clip latents}.{\BBCQ}
\newblock
\APACjournalVolNumPages{arXiv preprint arXiv:2204.06125}{1}{2}{3}.
\PrintBackRefs{\CurrentBib}

\bibitem [\protect \citeauthoryear {%
Rand%
, Arbesman%
\BCBL {}\ \BBA {} Christakis%
}{%
Rand%
\ \protect \BOthers {.}}{%
{\protect \APACyear {2011}}%
}]{%
rand2011dynamic}
\APACinsertmetastar {%
rand2011dynamic}%
\begin{APACrefauthors}%
Rand, D\BPBI G.%
, Arbesman, S.%
\BCBL {}\ \BBA {} Christakis, N\BPBI A.%
\end{APACrefauthors}%
\unskip\
\newblock
\APACrefYearMonthDay{2011}{}{}.
\newblock
{\BBOQ}\APACrefatitle {Dynamic social networks promote cooperation in experiments with humans} {Dynamic social networks promote cooperation in experiments with humans}.{\BBCQ}
\newblock
\APACjournalVolNumPages{Proceedings of the National Academy of Sciences}{108}{48}{19193--19198}.
\PrintBackRefs{\CurrentBib}

\bibitem [\protect \citeauthoryear {%
Shirado%
\ \BBA {} Christakis%
}{%
Shirado%
\ \BBA {} Christakis%
}{%
{\protect \APACyear {2017}}%
}]{%
shirado2017locally}
\APACinsertmetastar {%
shirado2017locally}%
\begin{APACrefauthors}%
Shirado, H.%
\BCBT {}\ \BBA {} Christakis, N\BPBI A.%
\end{APACrefauthors}%
\unskip\
\newblock
\APACrefYearMonthDay{2017}{}{}.
\newblock
{\BBOQ}\APACrefatitle {Locally noisy autonomous agents improve global human coordination in network experiments} {Locally noisy autonomous agents improve global human coordination in network experiments}.{\BBCQ}
\newblock
\APACjournalVolNumPages{Nature}{545}{7654}{370--374}.
\PrintBackRefs{\CurrentBib}

\bibitem [\protect \citeauthoryear {%
SonicRim%
}{%
SonicRim%
}{%
{\protect \APACyear {2001}}%
}]{%
sonicrim2001collective}
\APACinsertmetastar {%
sonicrim2001collective}%
\begin{APACrefauthors}%
SonicRim, L\BPBI S.%
\end{APACrefauthors}%
\unskip\
\newblock
\APACrefYearMonthDay{2001}{}{}.
\newblock
{\BBOQ}\APACrefatitle {Collective creativity} {Collective creativity}.{\BBCQ}
\newblock
\APACjournalVolNumPages{Design}{6}{3}{1--6}.
\PrintBackRefs{\CurrentBib}

\bibitem [\protect \citeauthoryear {%
Torrance%
}{%
Torrance%
}{%
{\protect \APACyear {1966}}%
}]{%
torrance1966torrance}
\APACinsertmetastar {%
torrance1966torrance}%
\begin{APACrefauthors}%
Torrance, E\BPBI P.%
\end{APACrefauthors}%
\unskip\
\newblock
\APACrefYearMonthDay{1966}{}{}.
\newblock
{\BBOQ}\APACrefatitle {Torrance tests of creative thinking} {Torrance tests of creative thinking}.{\BBCQ}
\newblock
\APACjournalVolNumPages{Educational and psychological measurement}{}{}{}.
\PrintBackRefs{\CurrentBib}

\bibitem [\protect \citeauthoryear {%
Tsvetkova%
, Yasseri%
, Pescetelli%
\BCBL {}\ \BBA {} Werner%
}{%
Tsvetkova%
\ \protect \BOthers {.}}{%
{\protect \APACyear {2024}}%
}]{%
tsvetkova2024new}
\APACinsertmetastar {%
tsvetkova2024new}%
\begin{APACrefauthors}%
Tsvetkova, M.%
, Yasseri, T.%
, Pescetelli, N.%
\BCBL {}\ \BBA {} Werner, T.%
\end{APACrefauthors}%
\unskip\
\newblock
\APACrefYearMonthDay{2024}{}{}.
\newblock
{\BBOQ}\APACrefatitle {A new sociology of humans and machines} {A new sociology of humans and machines}.{\BBCQ}
\newblock
\APACjournalVolNumPages{Nature Human Behaviour}{8}{10}{1864--1876}.
\PrintBackRefs{\CurrentBib}

\bibitem [\protect \citeauthoryear {%
Vaccaro%
, Almaatouq%
\BCBL {}\ \BBA {} Malone%
}{%
Vaccaro%
\ \protect \BOthers {.}}{%
{\protect \APACyear {2024}}%
}]{%
vaccaro2024combinations}
\APACinsertmetastar {%
vaccaro2024combinations}%
\begin{APACrefauthors}%
Vaccaro, M.%
, Almaatouq, A.%
\BCBL {}\ \BBA {} Malone, T.%
\end{APACrefauthors}%
\unskip\
\newblock
\APACrefYearMonthDay{2024}{}{}.
\newblock
{\BBOQ}\APACrefatitle {When combinations of humans and AI are useful: A systematic review and meta-analysis} {When combinations of humans and ai are useful: A systematic review and meta-analysis}.{\BBCQ}
\newblock
\APACjournalVolNumPages{Nature Human Behaviour}{}{}{1--11}.
\PrintBackRefs{\CurrentBib}

\bibitem [\protect \citeauthoryear {%
Wang%
\ \protect \BOthers {.}}{%
Wang%
\ \protect \BOthers {.}}{%
{\protect \APACyear {2020}}%
}]{%
wang2020minilm}
\APACinsertmetastar {%
wang2020minilm}%
\begin{APACrefauthors}%
Wang, W.%
, Wei, F.%
, Dong, L.%
, Bao, H.%
, Yang, N.%
\BCBL {}\ \BBA {} Zhou, M.%
\end{APACrefauthors}%
\unskip\
\newblock
\APACrefYearMonthDay{2020}{}{}.
\newblock
{\BBOQ}\APACrefatitle {Minilm: Deep self-attention distillation for task-agnostic compression of pre-trained transformers} {Minilm: Deep self-attention distillation for task-agnostic compression of pre-trained transformers}.{\BBCQ}
\newblock
\APACjournalVolNumPages{Advances in Neural Information Processing Systems}{33}{}{5776--5788}.
\PrintBackRefs{\CurrentBib}

\bibitem [\protect \citeauthoryear {%
Wiggins%
\ \BBA {} Forth%
}{%
Wiggins%
\ \BBA {} Forth%
}{%
{\protect \APACyear {2018}}%
}]{%
wiggins2018computational}
\APACinsertmetastar {%
wiggins2018computational}%
\begin{APACrefauthors}%
Wiggins, G\BPBI A.%
\BCBT {}\ \BBA {} Forth, J.%
\end{APACrefauthors}%
\unskip\
\newblock
\APACrefYearMonthDay{2018}{}{}.
\newblock
{\BBOQ}\APACrefatitle {Computational creativity and live algorithms} {Computational creativity and live algorithms}.{\BBCQ}
\newblock

\PrintBackRefs{\CurrentBib}

\bibitem [\protect \citeauthoryear {%
Zhang%
, Liao%
\BCBL {}\ \BBA {} Bellamy%
}{%
Zhang%
\ \protect \BOthers {.}}{%
{\protect \APACyear {2020}}%
}]{%
zhang2020effect}
\APACinsertmetastar {%
zhang2020effect}%
\begin{APACrefauthors}%
Zhang, Y.%
, Liao, Q\BPBI V.%
\BCBL {}\ \BBA {} Bellamy, R\BPBI K.%
\end{APACrefauthors}%
\unskip\
\newblock
\APACrefYearMonthDay{2020}{}{}.
\newblock
{\BBOQ}\APACrefatitle {Effect of confidence and explanation on accuracy and trust calibration in AI-assisted decision making} {Effect of confidence and explanation on accuracy and trust calibration in ai-assisted decision making}.{\BBCQ}
\newblock
\BIn{} \APACrefbtitle {Proceedings of the 2020 conference on fairness, accountability, and transparency} {Proceedings of the 2020 conference on fairness, accountability, and transparency}\ (\BPGS\ 295--305).
\PrintBackRefs{\CurrentBib}

\end{thebibliography}

\end{document}